\documentclass[twocolumn]{IEEEtran}
\usepackage{cite,algorithmic,array,url}
\usepackage{graphicx}
\usepackage[percent]{overpic}
\usepackage{amsfonts,times,amsmath,amssymb,amsthm}

\ifCLASSOPTIONcompsoc
\else
\fi
\ifCLASSOPTIONcaptionsoff
\let\MYoriglatexcaption\caption
 \renewcommand{\caption}[2][\relax]{\MYoriglatexcaption[#2]{#2}}
\fi
\theoremstyle{plain}

\newtheorem{lemma}{Lemma}
\newtheorem{proposition}{Proposition}
\newtheorem{corollary}{Corollary}

\theoremstyle{definition}
\newtheorem{definition}{Definition}

\theoremstyle{remark}
\newtheorem*{remark}{Remark}

\begin{document}

\title{The Cauchy-Schwarz divergence for Poisson point processes}
\author{Hung~Gia~Hoang, Ba-Ngu~Vo, Ba-Tuong~Vo, and~Ronald~Mahler 
\thanks{
Acknowledgement: The work of B.-N. Vo and B.-T. Vo are supported by the Australian Research Council under Future Fellowship FT0991854 and Discovery Early Career Research Award DE120102388 respectively.}
\thanks{H.~G. Hoang, B.-N. Vo, and B.-T. Vo are with the Department of Electrical
and Computer Engineering, Curtin University, Bentley, WA 6102, Australia
(email: \{hung.hoang,ba-ngu.vo,ba-tuong.vo\}@curtin.edu.au).} \thanks{%
R.~Mahler is with Random Sets LLC (email: mahlerronald@comcast.net).}
\thanks{Part of the paper has been presented at the 2014 IEEE Workshop on
Statistical Signal Processing, Gold Coast, Australia \cite{HVVM14}.}}
\maketitle
\IEEEpeerreviewmaketitle
\begin{abstract}
In this paper, we extend the notion of Cauchy-Schwarz divergence to point
processes and establish that the Cauchy-Schwarz divergence between the
probability densities of two Poisson point processes is half the squared $%
{L^{2}}$-distance between their intensity functions. Extension of
this result to mixtures of Poisson point processes and, in the case where
the intensity functions are Gaussian mixtures, closed form expressions for
the Cauchy-Schwarz divergence are presented. Our result also implies that
the Bhattacharyya distance between the probability distributions of two
Poisson point processes is equal to the square of the Hellinger distance
between their intensity measures. We illustrate the result via a sensor
management application where the system states are modeled as point
processes.
\end{abstract}





\section{Introduction}

The Poisson point process, which models ``no interaction'' or ``complete
spatial randomness'' in spatial point patterns, is arguably one of the best
known and most tractable of point processes \cite{DJ88, SKM95, kingman93poisson, van2000markov, MW04}. Point process theory is the
study of random counting measures with applications spanning numerous
disciplines, see for example \cite{DJ88, SKM95, MW04, Baddeley2007spatial, Baccelli10}. The Poisson point process itself arises
in forestry \cite{SP00forestry}, geology \cite{Oga99seismicity}, biology
\cite{Mar05neural}, particle physics \cite{STT81positron}, communication
networks \cite{Baccelli97, Haen05network, Haenggi09} and
signal processing \cite{MahlerPHD, Singhetal09, Caronetal11}. The
role of the Poisson point process in point process theory, in most respects,
is analogous to that of the normal distribution in random vectors \cite{CI80}.

Similarity measures between random variables are fundamental in information
theory and statistical analysis \cite{Cover}. Information theoretic
divergences, for example Kullback-Leibler, R\'{e}nyi (or $\alpha $%
-divergence) and their generalization Csisz\'{a}r-Morimoto (or Ali-Silvey),
Jensen-R\'{e}nyi, Cauchy-Schwarz etc., measure the difference between the
information content of the random variables. Similarity between random
variables can also be measured via the statistical distance between their probability
distributions, for example total variation, Bhattacharyya, Hellinger/Matusita, Wasserstein, etc. Some distances are actually special cases of $f$-divergences \cite{AliSilvey66}. Note that statistical distances are not necessarily proper metrics.

For point processes or random finite sets, similarity measures have been studied extensively in various application areas such as sensor management \cite{Sch93, Mah98senman, SKV07, HKB08, RVC11, HGH12} and neuroscience \cite{Victor05}.
However, so far except for trivial special cases, these similarity measures cannot be computed
analytically and require expensive approximations such as Monte Carlo.

In this paper, we present results on similarity measures for Poisson point
processes via the Cauchy-Schwarz divergence and its relationship to the
Bhattacharyya and Hellinger distances. In particular, we show that the
Cauchy-Schwarz divergence between two Poisson point processes is given by
the square of the $L^{2}$-distance between their intensity functions.
Geometrically, this result relates the angle subtended by the probability
densities of the Poisson point processes to the $L^{2}$-distance between
their corresponding intensity functions. For Gaussian mixture intensity
functions, their $L^{2}$-distance, and hence the Cauchy-Schwarz divergence
can be evaluated analytically. We also extend the result to the
Cauchy-Schwarz divergence for mixtures of Poisson point processes. In
addition, using our result on the Cauchy-Schwarz divergence, we show that
the Bhattacharyya distance between the probability distributions of two
Poisson point processes is the square of the Hellinger distance between
their respective intensity measures. The Poisson point process enjoys a
number of nice properties \cite{DJ88, SKM95, kingman93poisson}, and our results are useful additions. We illustrate the use of our result
on the Cauchy-Schwarz divergence in a sensor management application for
multi-target tracking involving the Probability Hypothesis Density (PHD)
filter \cite{MahlerPHD}.

The organization of the paper is as follows. Background on point processes
and the Cauchy-Schwarz divergence is provided in Section~\ref{back}. Section~%
\ref{main} presents the main results of the paper that establish the
analytical formulation for the Cauchy-Schwarz divergence and Bhattacharyya
distance between two Poisson point processes. In Section~\ref{sencon}, the
application of the Cauchy-Schwarz divergence to sensor management, including
numerical examples, is studied. Finally, Section~\ref{sum} concludes the
paper.

\section{Background}

\label{back}

In this work we consider a state space $\mathcal{X}\subseteq \mathbb{R}^{d}$%
, and adopt the inner product notation $\left\langle
f,g\right\rangle\triangleq \int f(x)g(x)dx$; the $L^{2}$-norm notation $%
\left\Vert f \right\Vert \triangleq \sqrt{\left\langle f,f\right\rangle }$;
the multi-target exponential notation $h^{X}\triangleq
\prod_{x\in X}h(x)$, where $h$ is a real-valued function, with $h^{\emptyset
}=1$ by convention; and the indicator function notation
\begin{equation*}
1_{B}(x)\triangleq \left\{
\begin{array}{l}
1,\text{ if }x\in B \\
0,\text{ otherwise}%
\end{array}%
\right. .
\end{equation*}
The notation $\mathcal{N}(x;m , Q)$ is used to explicitly denote the probability density of a Gaussian random variable with mean $m$ and covariance $Q$, evaluated at $x$.

\subsection{Point processes}

\label{point_pro}

This section briefly summarizes concepts in point process theory needed for
the exposition of our result. Point process theory, in general, is concerned
with \emph{random counting measures}. Our result is restricted to
simple-finite point processes, which can be regarded as \emph{random finite
sets}. For simplicity, we omit the prefix ``simple-finite'' in the rest of
the paper. For an introduction to the subject we refer the reader to the
article \cite{Baddeley2007spatial}, and for detailed treatments, books
such as \cite{DJ88, SKM95, van2000markov, MW04}.

A \emph{point process} or \emph{random finite set} $X$ on $\mathcal{X}$
is a random variable taking values in $\mathcal{F}(\mathcal{X})$, the space of
finite subsets of $\mathcal{X}$. Let $|X|$ denotes the number of elements in
a set $X$. A point process $X$ on $\mathcal{X}$ is said to be \emph{Poisson}
with a given \emph{intensity function} $u$ (defined on $\mathcal{X}$) if
\cite{DJ88, SKM95}:

\begin{enumerate}
\item for any $B\subseteq \mathcal{X}$ such that $\left\langle
u,1_{B}\right\rangle <\infty $, the random variable $|X\cap B|$ is Poisson
distributed with mean $\left\langle u,1_{B}\right\rangle $,

\item for any disjoint $B_{1},...,B_{i}\subseteq \mathcal{X}$, the random
variables $|X\cap B_{1}|,...,|X\cap B_{i}|$ are independent.
\end{enumerate}

Since $\left\langle u,1_{B}\right\rangle $ is the expected number of points
of $X$ in the region $B$, the intensity value $u(x)$ can be interpreted as
the instantaneous expected number of points per unit hyper-volume at $x$.
Consequently, $u(x)$ is not dimensionless in general. If hyper-volume (on $%
\mathcal{X}$) is measured in units of $K$ (e.g. $m^{d}$, $cm^{d}$, in$^{d}$,
etc.) then the intensity function $u$ has unit $K^{-1}$.

The number of points of a Poisson point process $X$ is Poisson distributed
with mean $\left\langle u,1\right\rangle $, and conditional on the number of
points the elements $x$ of $X$, are independently and identically distributed
(i.i.d.) according to the probability density $u(\cdot )/\left\langle
u,1\right\rangle $ \cite{DJ88, SKM95, van2000markov, MW04}. It is implicit
that $\left\langle u,1\right\rangle $ is finite since we only consider
simple-finite point processes.

The probability distribution of a Poisson point process $X$ with intensity
function $u$ is given by \cite[pp. 15]{MW04}
\begin{multline}
\!\!\!\Pr (X\in \mathcal{T})=\\ \sum_{i=0}^{\infty }\!\frac{e^{-\left\langle u,1\right\rangle }}{i{!}}\!\int_{%
\mathcal{X}^{i}}\!1_{\mathcal{T}}(\{x_{1},...,x_{i}\})u^{\{x_{1},...,x_{i}%
\}}d(x_{1},...,x_{i}),  \label{eq:Poissonmeas}
\end{multline}%
for any (measurable) subset $\mathcal{T}$ of $\mathcal{F}(\mathcal{X})$,
where $\mathcal{X}^{i}$ denotes the $i^{\text{th}}$-fold Cartesian product
of $\mathcal{X}$, with the convention $\mathcal{X}^{0}=\{\emptyset \}$, and
the integral over $\mathcal{X}^{0}$ is $1_{\mathcal{T}}(\emptyset )$. A
Poisson point process is completely characterized by its intensity function
(or more generally the intensity measure).

Probability densities of point processes considered in this work are defined
with respect to the reference measure $\mu$ given by%
\begin{equation}
\mu (\mathcal{T})=\sum_{i=0}^{\infty }\frac{1}{i{!K}^{i}}\int_{\mathcal{X}%
^{i}}1_{\mathcal{T}}(\{x_{1},...,x_{i}\})d(x_{1},...,x_{i})
\label{eq:referencemeasure}
\end{equation}%
for any (measurable) subset $\mathcal{T}$ of $\mathcal{F}(\mathcal{X})$. The
measure $\mu $ is analogous to the Lebesque measure on $\mathcal{X}$ (indeed
it is the unnormalized distribution of a Poisson point process with unit
intensity $u=1/K$ when the state space $\mathcal{X}$ is bounded). Moreover,
it was shown in \cite{VSD05} that for this choice of reference measure, the
integral of a function $f:\mathcal{F}(\mathcal{X})\rightarrow \mathbb{R}$,
given by%
\begin{equation}
\int \!\!f(X)\mu (dX)=\sum_{i=0}^{\infty }\frac{1}{i{!K}^{i}}\!\!\int_{%
\mathcal{X}^{i}}\!f(\{x_{1},...,x_{i}\})d(x_{1},...,x_{i}),
\label{eq:integral}
\end{equation}%
is equivalent to Mahler's set integral \cite{Mah07}. Note that the reference
measure $\mu $, and the integrand $f$ are all dimensionless.

Our main result involves Poisson point processes with probability densities
of the form%
\begin{equation}
f(X)=e^{-\left\langle u,1\right\rangle }\left[ Ku\right] ^{X}.
\label{eq:Poissondensity}
\end{equation}%
Note that for any (measurable) subset $\mathcal{T}$ of $\mathcal{F}(\mathcal{%
X})$%
\begin{eqnarray*}
&&\!\!\!\!\!\!\!\!\!\!\!\!\!\!\!\!\!\!\!\int_{\mathcal{T}}f(X)\mu (dX) \\
&=&\!\!\!\int 1_{\mathcal{T}}(X)f(X)\mu (dX) \\
&=&\!\!\!\sum_{i=0}^{\infty }\!\frac{e^{-\left\langle u,\!1\right\rangle }}{i%
{!}}\!\!\int_{\mathcal{X}^{i}}\!\!1_{\mathcal{T}}(\{x_{1},...,x_{i}\})u^{\!%
\{x_{1},...,x_{i}\}\!}d(x_{1},...,x_{i}).
\end{eqnarray*}%
Thus, comparing with \eqref{eq:Poissonmeas}, $f$ is indeed a probability
density (with respect to $\mu $) of a Poisson point process with intensity
function $u$.

\subsection{The Cauchy-Schwarz divergence}

The Cauchy-Schwarz divergence is based on the Cauchy-Schwarz inequality for
inner products, and is defined for two random vectors with probability
densities $f$ and $g$ by \cite{KHP11CFCSdiv}
\begin{equation}
D_{CS}(f,g)=-\ln \frac{\left\langle f,g\right\rangle }{\left\Vert
f\right\Vert \left\Vert g\right\Vert }.  \label{so_CS_div}
\end{equation}%
The argument of the logarithm in \eqref{so_CS_div} is non-negative (since
probability densities are non-negative) and does not exceed one (by the
Cauchy-Schwarz inequality). Moreover, this quantity can be interpreted as the cosine of
the angle subtended by $f$ and $g$ in $L^{2}(\mathcal{X},\mathbb{R})$, the
space of square integrable functions taking $\mathcal{X}$ to $\mathbb{R}$.
Note that $D_{CS}(f,g)$ is symmetric and positive unless $f=g$, in which
case $D_{CS}(f,g)=0$.

Geometrically, the Cauchy-Schwarz divergence determines the information
\textquotedblleft difference\textquotedblright\ between random vectors from
the angle between their probability densities. The Cauchy-Schwarz divergence
can also be interpreted as an approximation to the Kullback-Leibler
divergence \cite{KHP11CFCSdiv}. While the Kullback-Leibler divergence can be
evaluated analytically for Gaussians (random vectors) \cite{KL51, Lin91},
for the more versatile class of Gaussian mixtures, only Jensen-R\'{e}nyi and
Cauchy-Schwarz divergences can be evaluated in closed form \cite%
{Wang09JensenRenyi,KHP11CFCSdiv}. Hence, the Cauchy-Schwarz
divergence between two densities of random variables has been employed in
many information theoretic applications, especially in machine learning and
pattern recognition \cite{Jen05CSdiv,VHS05,JPE06,KHP11CFCSdiv,HGP11}.

\section{The Cauchy-Schwarz divergence for Poisson point processes}

\label{main}

This section presents the main theoretical results of the paper. Subsection~%
\ref{CS_for_PPP} establishes the Cauchy-Schwarz divergence for general
Poisson point processes. Subsection~\ref{GMI} presents analytical solution
for Poisson point processes with Gaussian mixture intensities while
subsection~\ref{Poiss_mix} details the solution for mixtures of Poisson
point processes. Finally, subsection~\ref{Bhatt} presents a result on the
Bhattacharyya distance between two Poisson processes.

\subsection{Cauchy-Schwarz divergence for Poisson point processes}

\label{CS_for_PPP}

For point processes, the Csisz\'{a}r-Morimoto divergence, which includes the
Kullback-Leibler and R\'{e}nyi, were formulated in \cite{Mah98senman} by
replacing the standard (Lebesque) integral with the set integral which is
defined for a Finite Set Statistics (FISST) density $\phi $ as follows \cite{Mah07}
\begin{equation*}
\int \phi (X)\delta {X}=\sum_{i=0}^{\infty }\frac{1}{i!}\int \phi
(\{x_{1},...,x_{i}\})d(x_{1},...,x_{i}).
\end{equation*}%
The FISST density $\phi $ is not a probability density, but is closely
related to a probability density, see \cite{VSD05} for further details. Note
that $\phi (\{x_{1},...,x_{i}\})$ has unit $K^{-i}$, since the infinitesimal
hyper-volume $d(x_{1},...,x_{i})$ has unit $K^{i}$. Thus, $\phi (X)$ has
different units for different cardinalities of $X$.

Unlike the Csisz\'{a}r-Morimoto divergence, the Cauchy-Schwarz divergence,
however, cannot be extended to point processes by simply replacing the
standard integral with the set integral. To see this, consider the na\"{\i}ve
inner product between two FISST densities $\phi $ and $\varphi $ via the set
integral:
\begin{align*}
\left\langle \phi ,\varphi \right\rangle & =\int \phi (X)\varphi (X)\delta {X%
}\\
& =\sum_{i=0}^{\infty }\frac{1}{i!}\int \phi (\{x_{1},...,x_{i}\})\varphi
(\{x_{1},...,x_{i}\})d(x_{1},...,x_{i});
\end{align*}
since the $i$-th term in the above sum has units of $K^{-i}$, the sum itself is meaningless because the terms cannot be added together due to unit mismatch, e.g. if $K=m^{3}$, then the first term is unitless, the second term is in $%
m^{-3}$, the third term is in $m^{-6}$, etc. Indeed such na\"{\i}ve inner product has been used incorrectly in \cite{DHJ12}.

Using the standard notion of density and integration summarized in
subsection~\ref{point_pro}, we can define the inner product 
\[\left\langle
f,g\right\rangle_{\!\mu }=\int f(X)g(X)\mu (dX),\] 
and corresponding norm 
\[\left\Vert f\right\Vert _{\mu }\triangleq \sqrt{\left\langle
f,f\right\rangle_{\!\mu }}\]
 on $L^{2}(\mathcal{F}(\mathcal{X}),\mathbb{R})$.
Such forms for the inner product and norm are well-defined because the
densities $f$, $g$ and reference measure $\mu $ are all unitless.

Interestingly, the inner product between multi-object exponentials is given
by the following result.

\begin{lemma}
\label{CSmultobj} Let $f(X)=r^{X}$ and $g(X)=s^{X}$ with $r$, $s\in L^{2}(%
\mathcal{X},\mathbb{R})$. Then $\left\langle f,g\right\rangle_{\!\mu }=$ $%
e^{K^{-1}\left\langle r,s\right\rangle }$.
\end{lemma}

\textit{Proof:}
\allowdisplaybreaks{
\begin{align*}
\left\langle f,g\right\rangle_{\!\mu }& =\int \left[ rs\right] ^{X}\mu (dX) \\
& =\sum_{i=0}^{\infty }\frac{1}{i{!}K^{i}}\left[ \int_{\mathcal{X}}r(x)s(x)dx%
\right] ^{i}\text{ \ (using \eqref{eq:integral})} \\
& =\sum_{i=0}^{\infty }\frac{\left\langle r,s\right\rangle ^{i}}{i!K^{i}}%
=e^{K^{-1}\left\langle r,s\right\rangle}.\QED
\end{align*}}

In the spirit of using the angle between probability densities to determine
the information \textquotedblleft difference\textquotedblright , the
Cauchy-Schwarz divergence can be extended to point processes as follows.

\begin{definition}
\label{CSdef} The Cauchy-Schwarz divergence between the probability
densities $f$ and $g$ of two point processes with respect to the reference
measure $\mu $ is defined by%
\begin{equation}
D_{CS}(f,g)=-\ln \frac{\left\langle f,g\right\rangle_{\!\mu }}{\left\Vert
f\right\Vert _{\mu }\left\Vert g\right\Vert _{\mu }}.  \label{mo_CS_div}
\end{equation}
\end{definition}

The above definition of the Cauchy-Schwarz divergence can be equivalently
expressed in terms of set integrals as follows. Let $\phi $ and $\varphi $
denote the FISST densities of the respective point processes. Using the
relationship between the FISST density and the Radon-Nikodym derivative in
\cite{VSD05}, the corresponding probability densities relative to $\mu $ are
given by $f(X)=K^{|X|}\phi (X)$ and $g(X)=K^{|X|}\varphi (X)$. Since%
\begin{align*}
\left\langle f,g\right\rangle_{\!\mu }& =\sum_{i=0}^{\infty }\frac{1}{i{!}%
K^{i}}\!\!\int_{\mathcal{X}^{i}}\!K^{i}\phi(\{x_{1},...,x_{i}\}) \times\\
&\qquad\qquad\qquad\;\; K^{i}\varphi(\{x_{1},...,x_{i}\})d(x_{1},...,x_{i}) \\
& =\int K^{|X|}\phi (X)\varphi (X)\delta {X}
\end{align*}
the Cauchy-Schwarz divergence can be written as
\begin{equation*}
D_{CS}(\phi ,\varphi )=-\ln \frac{\int K^{|X|}\phi (X)\varphi (X)\delta {X}}{%
\sqrt{\int K^{|X|}\phi ^{2}(X)\delta {X}\int K^{|X|}\varphi ^{2}(X)\delta {X}%
}}.
\end{equation*}

The following proposition asserts that \emph{the Cauchy-Schwarz divergence
between two Poisson point processes is half the squared distance between
their intensity functions}.

\begin{proposition}
\label{CSpro} The Cauchy-Schwarz divergence between the probability
densities $f$ and $g$ of two Poisson point processes with respective
intensity functions $u$ and $v\in L^{2}(\mathcal{X},\mathbb{R})$ (measured
in units of $K^{-1}$), is given by
\begin{equation}
D_{CS}(f,g)=\frac{K}{2}\left\Vert u-v\right\Vert ^{2}.  \label{DCS3}
\end{equation}
\end{proposition}

\textit{Proof:} Substituting $f(X)=e^{-\left\langle u,1\right\rangle }\left[
Ku\right] ^{X}$, $g(X)=e^{-\left\langle v,1\right\rangle }\left[ Kv\right]
^{X} $ into \eqref{mo_CS_div} and canceling out the constants $e^{-\langle
u,1\rangle}$, $e^{-\langle v,1\rangle}$ we have
\begin{equation*}
D_{CS}(f,g)=-\ln\left(\frac{\left\langle [Ku]^{(\cdot)},[Kv]^{(\cdot)}\right\rangle_{\!\mu}} {%
\left\langle [Ku]^{(\cdot)},[Ku]^{(\cdot)}\right\rangle_{\!\mu}^{\frac{1}{2}}\left\langle
[Kv]^{(\cdot)},[Kv]^{(\cdot)}\right\rangle_{\!\mu}^{\frac{1}{2}}}\right)
\end{equation*}
Applying Lemma~\ref{CSmultobj} to the above equation gives
\begin{align*}
D_{CS}(f,g)& =-\ln \left( e^{K\left\langle u,v\right\rangle -\frac{K}{2}%
\left\langle u,u\right\rangle -\frac{K}{2}\left\langle v,v\right\rangle
}\right) \\
& =-\ln\left(e^{-\frac{K}{2}\left(\left\langle u,u\right\rangle
-2\left\langle u,v\right\rangle +\left\langle v,v\right\rangle\right)}
\right) \\
& =\frac{K}{2}\left\Vert u-v\right\Vert ^{2}.\QED
\end{align*}

\begin{figure}[htb]
\centering
\begin{overpic}[scale=0.5]{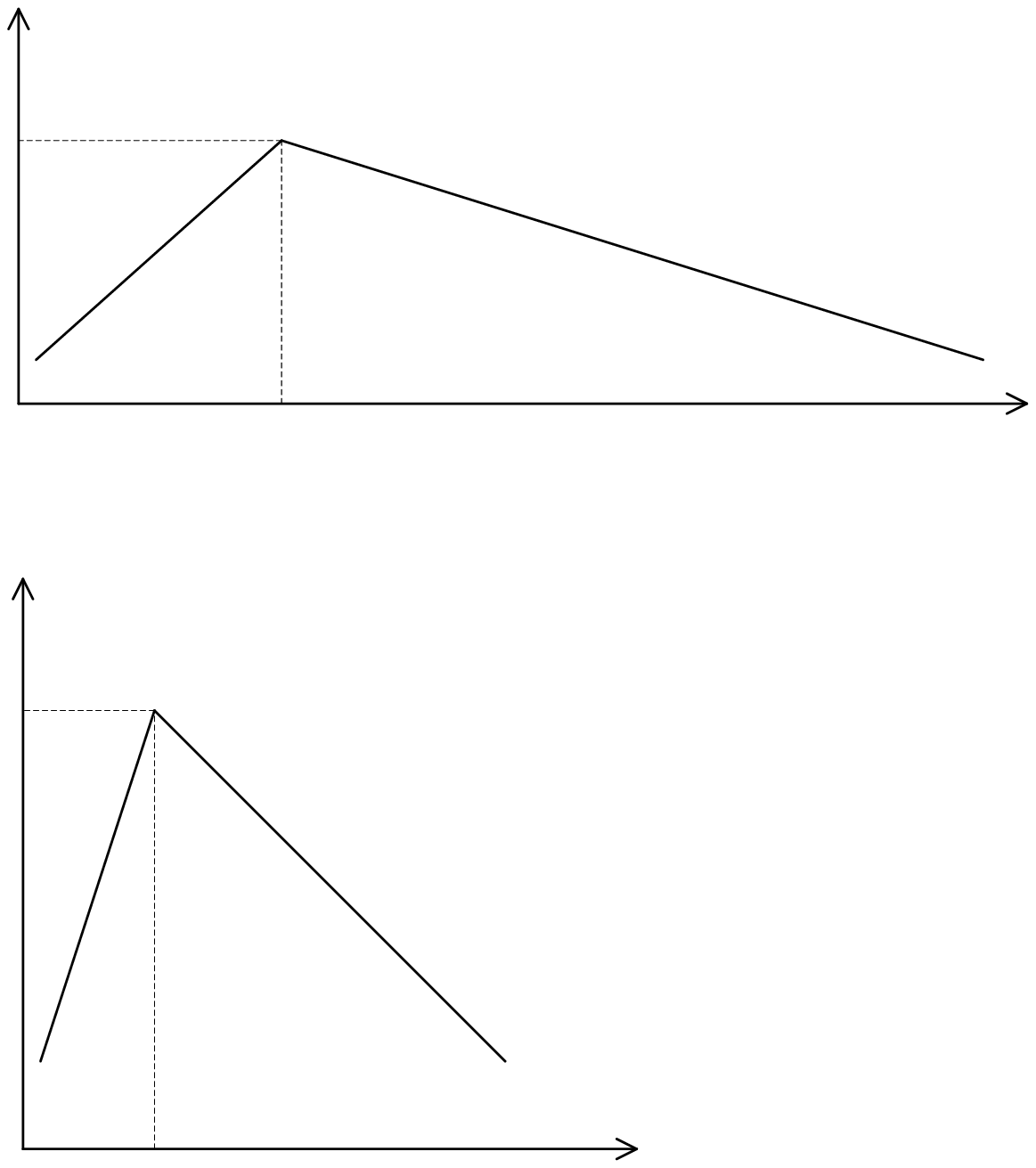}
		\footnotesize
		\put(13,92){$u$ [${K}^{-1}$]}
    \put(6,83){${u_0}$}
		\put(30,60){${x}_0$}
		\put(86,60){${x} [K]$}
		\put(13,49){$u^{\;\prime}$ [$\rho^{-1}K^{-1}$]}
		\put(3.5,40){$\rho{u_0}$}
		\put(57,4){${x^{\;\prime}}$ [$\rho{}K$]}
		\put(16,3){$\rho^{-1}{x}_0$}
\end{overpic}
\caption{Change of unit in the state space}
\label{unit}
\end{figure}

Note that since the intensity functions have units of $K^{-1}$, $\left\Vert
u-v\right\Vert ^{2}$ also has unit of $K^{-1}$ and hence $K\left\Vert
u-v\right\Vert ^{2}$ is unitless. Moreover, $K\left\Vert u-v\right\Vert ^{2}$%
, referred to as the \emph{squared distance} between the intensity functions
$u$ and $v$, takes on the same value regardless of the choice of measurement
unit. Suppose that the unit of the hyper-volume in the state space $\mathcal{%
X}$ has been changed from $K$ to $\rho {K}$ (for example, from $dm^{3}$ to $%
m^{3}=10^{3}dm^{3}$) as illustrated in Fig.~\ref{unit}. The change of unit
inevitably leads to the change in numerical values of the two intensity
functions (for example, the intensity measured in $m^{-3}$, which is the
expected number of points per cubic meter, is one thousand times the
intensity measured in $dm^{-3}$). However, these changes cancel each other
in the product $\frac{\rho {K}}{2}\left\Vert u_{\rho }-v_{\rho }\right\Vert
^{2}$ such that the \emph{squared distance} remains unchanged.

\begin{figure}[!htb]
\centering
\begin{overpic}[scale=0.46]{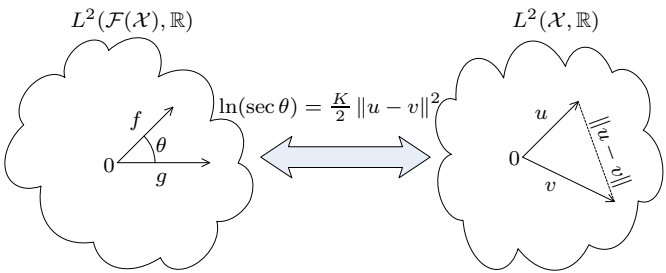}
		\footnotesize
    \put(10,37){$L^2(\mathcal{F}(\mathcal{X}),\mathbb{R})$}
		\put(77,37){$L^2(\mathcal{X},\mathbb{R})$}
		\put(23,18){$\theta$}
		\put(15,15){$0$}
		\put(76.5,16){$0$}
		\put(19,22){$f$}
		\put(23,13.5){$g$}
		\put(80.5,23){$u$}
		\put(82,12){$v$}
		\put(88.5,23){\rotatebox{290}{$\left\|u-v\right\|$}}
		\put(32.5,24){$\ln(\sec \theta) = \frac{K}{2}\left\|u-v\right\|^2$}
  \end{overpic}
\caption{Geometric interpretation of Proposition~\protect\ref{CSpro}}
\label{mapping}
\end{figure}

Proposition~\ref{CSpro} has a nice geometric interpretation that relates the angle
subtended by the probability densities\ in $L^{2}(\mathcal{F}(\mathcal{X}),%
\mathbb{R})$ to the distance between the corresponding intensity functions
in $L^{2}(\mathcal{X},\mathbb{R})$ as depicted in Fig.~\ref{mapping}. More
concisely:\emph{\ the secant of the angle between the probability densities
of two Poisson point processes equals the exponential of half the squared
distance between their intensity functions.}

The above result has important implications in the approximation of
Poisson point processes through their intensity functions. It is intuitive
that the \textquotedblleft difference\textquotedblright\ between the Poisson
distributions vanishes as the distance between their intensity functions
tends to zero. However, it was not clear that a reduction in the error
between the intensity functions necessarily implies a reduction in the
\textquotedblleft difference\textquotedblright\ between the corresponding
distributions. Our result not only verifies that the \textquotedblleft
difference\textquotedblright\ between the distributions is reduced, it also
quantifies the reduction.

\subsection{Gaussian Mixture Intensities}

\label{GMI}

In general, the $L^{2}$-distance between the intensity functions, and hence
the Cauchy-Schwarz divergence, cannot be numerically evaluated in closed
form. However, for Poisson point processes with Gaussian mixture intensity
functions, applying the following identity for Gaussian probability density functions \cite[pp. 200]{Ras05_Gauss}
\begin{equation*}
\left\langle \mathcal{N}(\cdot ;\mu _{0},\Sigma _{0}),\mathcal{N}(\cdot ;\mu
_{1},\Sigma _{1})\right\rangle =\mathcal{N}(\mu _{0};\mu _{1},\Sigma
_{0}+\Sigma _{1}),
\end{equation*}%
to \eqref{DCS3} yields an analytic expression for the Cauchy-Schwarz
divergence. This is stated more concisely in the following result.

\begin{corollary}
\label{CSdivGM} The Cauchy-Schwarz divergence between two Poisson point
processes with Gaussian mixture intensities:
\begin{subequations}
\label{GM}
\begin{align}
u(x)& =\sum_{i=1}^{N_{u}}w_{u}^{(i)}\mathcal{N}(x;m_{u}^{(i)},P_{u}^{(i)}),
\\
v(x)& =\sum_{i=1}^{N_{v}}w_{v}^{(i)}\mathcal{N}(x;m_{v}^{(i)},P_{v}^{(i)})
\end{align}%
(measured in units of $K^{-1}$) is given by
\end{subequations}
\begin{multline}
D_{CS}(f,g)= \\
\hfill\frac{1}{2}\sum_{i=1}^{N_{u}}\sum_{j=1}^{N_{u}}w_{u}^{(i)}w_{u}^{(j)}%
\mathcal{N}\left( m_{u}^{(i)};m_{u}^{(j)},P_{u}^{(i)}+P_{u}^{(j)}\right)+\\
\hfill\frac{1}{2}\sum_{i=1}^{N_{v}}\sum_{j=1}^{N_{v}}w_{v}^{(i)}w_{v}^{(j)}%
\mathcal{N}\left( m_{v}^{(i)};m_{v}^{(j)},P_{v}^{(i)}+P_{v}^{(j)}\right)-\\
\hfill\sum_{i=1}^{N_{u}}\sum_{j=1}^{N_{v}}w_{u}^{(i)}w_{v}^{(j)}\mathcal{N}\left(
m_{u}^{(i)};m_{v}^{(j)},P_{u}^{(i)}+P_{v}^{(j)}\right)  \label{clf_CS}
\end{multline}
\end{corollary}
In terms of computational complexity, each term in \eqref{clf_CS} involves evaluations of a Gaussian probability density function within a double sum. Hence, if we use the standard Gauss-Jordan elimination for matrix inversions, computing $D_{CS}(f,g)$ is quadratic in the number of Gaussian components and cubic in the state dimension (i.e. $O(N_v^2d^3)$, assuming $N_v\geq N_u$). The complexity can be reduced to $O(N_v^2d^{2.373})$ if the optimized Coppersmith-Winograd algorithm \cite{LeGall14} was employed in place of the Gauss-Jordan elimination.

This Corollary has important implications in Gaussian mixture reduction for
intensity functions. The result provides mathematical justification for
Gaussian mixture reduction for intensity functions based on $L^{2}$-error.
Furthermore, since Gaussian mixtures can approximate any density to any
desired accuracy \cite{Lo72}, Corollary~\ref{CSdivGM} enables the
Cauchy-Schwarz divergence between two Poisson point processes to be
approximated to any desired accuracy.

\subsection{Mixture of Poisson point processes}

\label{Poiss_mix}

Proposition~\ref{CSpro} can be easily extended to mixtures of Poisson point
processes, i.e. those whose probability densities can be written as a
weighted sum of Poisson point process densities:
\vspace{-0.5cm}
\begin{subequations}
\label{mixPP}
\begin{align}
f(X)& =\sum_{i=1}^{N_{f}}w_{f}^{(i)}e^{-\left\langle
u_{i},1\right\rangle }\left[K{u_{i}}\right]^{X}, \\
g(X)& =\sum_{i=1}^{N_{g}}w_{g}^{(i)}e^{-\left\langle
v_{i},1\right\rangle }\left[K{v_{i}}\right]^{X},
\end{align}%
\end{subequations}
where $\sum_{i=1}^{N_{f}}w_{f}^{(i)}=\sum_{i=1}^{N_{g}}w_{g}^{(i)}=1$. Such
point processes have applications in immunology \cite{JMK09}, neural data
analysis \cite{KB10}, criminology \cite{Tad10}, and machine learning \cite%
{PV14}.

Substituting \eqref{mixPP} into \eqref{mo_CS_div} and applying Lemma~\ref%
{CSmultobj}, the Cauchy-Schwarz divergence between two mixtures of Poisson
point processes is stated as follows.

\begin{corollary}
\label{CSmix} The Cauchy-Schwarz divergence between two mixtures of Poisson
point processes given in \eqref{mixPP} is

\vspace{-0.3cm}
\begin{align}\label{eq:DCSmixture}
D_{CS}(f,g) =&-\ln \left(\sum_{i=1}^{N_{f}}\sum_{j=1}^{N_{g}}w_{f}^{(i)}w_{g}^{(j)}\frac{e^{K\langle u_{i},v_{j}\rangle}}{e^{\langle u_i+v_j,1\rangle}}\right)+\notag\\
&\;\frac{1}{2}\ln \left(\sum_{i=1}^{N_{f}}\sum_{j=1}^{N_{f}}w_{f}^{(i)}w_{f}^{(j)}\frac{e^{K\langle u_{i},u_{j}\rangle}}{e^{\langle u_i+u_j,1\rangle}}\right)+\notag\\
&\;\frac{1}{2}\ln \left(\sum_{i=1}^{N_{g}}\sum_{j=1}^{N_{g}}w_{g}^{(i)}w_{g}^{(j)}\frac{e^{K\langle v_{i},v_{j}\rangle}}{e^{\langle v_i+ v_j,1\rangle}}\right).
\end{align}
Furthermore, if the intensity function of each Poisson point process
component is a Gaussian mixture (in units of $K^{-1}$):
\vspace{-0.1cm}
\begin{align*}
u_{i}(x)& =\sum_{\ell =1}^{N_{u_i}}\omega _{u_{i}}^{(\ell )}\mathcal{N}%
(x;m_{u_{i}}^{(\ell )},P_{u_{i}}^{(\ell )}), \\
v_{j}(x)& =\sum_{\ell =1}^{N_{v_j}}\omega _{v_{j}}^{(\ell )}\mathcal{N}%
(x;m_{v_{j}}^{(\ell )},P_{v_{j}}^{(\ell )}),
\end{align*}%
then $D_{CS}(f,g)$ can be evaluated analytically by substituting the
following equations into \eqref{eq:DCSmixture}
\vspace{-.1cm}
\allowdisplaybreaks{\begin{align*}
K\langle u_{i},u_{j}\rangle & =\sum_{\ell =1}^{N_{u_i}}\sum_{k
=1}^{N_{u_j}}\omega _{u_{i}}^{(\ell )}\omega _{u_{j}}^{(k )}\mathcal{N}%
\left( m_{u_{i}}^{(\ell )};m_{u_{j}}^{(k )},P_{u_{i}}^{(\ell
)}+P_{u_{j}}^{(k )}\right) , \\
K\langle v_{i},v_{j}\rangle & =\sum_{\ell =1}^{N_{v_i}}\sum_{k
=1}^{N_{v_j}}\omega _{v_{i}}^{(\ell )}\omega _{v_{j}}^{(k )}\mathcal{N}%
\left( m_{v_{i}}^{(\ell )};m_{v_{j}}^{(k )},P_{v_{i}}^{(\ell
)}+P_{v_{j}}^{(k )}\right) , \\
K\langle u_{i},v_{j}\rangle & =\sum_{\ell =1}^{N_{u_i}}\sum_{k
=1}^{N_{v_j}}\omega _{u_{i}}^{(\ell )}\omega _{v_{j}}^{(k )}\mathcal{N}%
\left( m_{u_{i}}^{(\ell )};m_{v_{j}}^{(k )},P_{u_{i}}^{(\ell
)}+P_{v_{j}}^{(k )}\right),\\
\langle u_i+v_j,1\rangle &=\sum_{\ell =1}^{N_{u_i}}\sum_{k
=1}^{N_{v_j}}\omega _{u_{i}}^{(\ell )}+\omega _{v_{j}}^{(k )},\\
\langle u_i+u_j,1\rangle &=\sum_{\ell =1}^{N_{u_i}}\sum_{k
=1}^{N_{u_j}}\omega _{u_{i}}^{(\ell )}+\omega _{u_{j}}^{(k )},\\
\langle v_i+v_j,1\rangle &=\sum_{\ell =1}^{N_{v_i}}\sum_{k
=1}^{N_{v_j}}\omega _{v_{i}}^{(\ell )}+\omega _{v_{j}}^{(k )}.
\end{align*}}
\end{corollary}

\subsection{Bhattacharyya distance for Poisson point processes}
\label{Bhatt}

The Cauchy-Schwarz divergence is based on the angle between two probability
densities (with respect to a reference measure), and is not necessarily
invariant to the choice of reference measure. Closely related to the
Cauchy-Schwarz divergence is the Bhattacharyya distance between two
probability measures \cite{Bha43}.

\begin{definition}
\label{BDdef} The Bhattacharyya distance between to probability measures $F$
and $G$, is defined by
\begin{equation}
D_{B}(F,G)=-\ln \left\langle \sqrt{\frac{dF}{d\mu }},\sqrt{\frac{dG}{d\mu }}%
\right\rangle _{\mu }
\end{equation}
where $\mu $ is any measure dominating $F$ and $G$. The inner product in the
above definition, denoted by $C_{B}(F,G)$, is called the Bhattacharyya coefficient and is invariant to the choice of
reference measure $\mu $ \cite{Bha43}. 
\end{definition}

Unlike the Cauchy-Schwarz divergence, the Bhattacharyya distance avoids the requirement of square
integrable probability densities since square roots of probability
densities are always square integrable. Note also that the Bhattacharyya distance can be expressed as the
Cauchy-Schwarz divergence between the square roots of the probability
densities, i.e. for any $\mu $ that dominates $F$ and $G$%
\begin{equation}
D_{B}(F,G)=D_{CS}\left( \sqrt{\frac{dF}{d\mu }},\sqrt{\frac{dG}{d\mu }}%
\right)
\end{equation}%
Hence, Proposition \ref{CSpro} can be applied to relate the Bhattacharyya
distance between the probability distributions of Poisson point processes to
their intensity functions.

\begin{corollary}
\label{BDpro} The Bhattacharyya distance between the probability
distributions $F$ and $G$ of two Poisson point processes with respective
intensity measures $U$ and $V$ (assumed to have densities with respect to
the Lebesque measure), is given by%
\begin{equation}
D_{B}(F,G)=D_{H}^{2}(U,V),  \label{BDeq}
\end{equation}%
where%
\begin{equation*}
D_{H}(U,V)=\frac{1}{\sqrt{2}}\left\Vert \sqrt{\frac{dU}{d\lambda }}-\sqrt{%
\frac{dV}{d\lambda }}\right\Vert ,
\end{equation*}
is the Hellinger distance between the measures $U,$ and $V$, (which is
invariant to the choice of reference measure).
\end{corollary}

\textit{Proof:} Let $u$ and $v$ be densities (measured in units of $K^{-1}$)
of $U$ and $V$ relative to the Lebesque measure $\lambda $. Then the
densities of $F$ and $G$ relative to $\mu $, are given by $%
f(X)=e^{-\left\langle u,1\right\rangle }\left[ Ku\right] ^{X}$, $%
g(X)=e^{-\left\langle v,1\right\rangle }\left[ Kv\right] ^{X}$. From
Proposition \ref{CSpro} the Cauchy-Schwarz divergence between $\sqrt{f(X)}%
\propto \left[ K\sqrt{u/K}\right] ^{X}$, and $\sqrt{g(X)}\propto \left[ K%
\sqrt{v/K}\right] ^{X}$ is given by%
\begin{align*}
D_{CS}\left( \sqrt{f},\sqrt{g}\right) & =\frac{K}{2}\left\Vert \sqrt{\frac{u%
}{K}}-\sqrt{\frac{v}{K}}\right\Vert ^{2}, \\
& =\frac{1}{2}\left\Vert \sqrt{u}-\sqrt{v}\right\Vert ^{2}, \\
& =D_{H}^{2}(U,V).\QED
\end{align*}

The above Corollary asserts that \emph{the Bhattacharyya distance between
two Poisson point processes is the squared Hellinger distance between their
intensity measures}. Moreover, the square of the Hellinger distance can be
expanded as
\begin{eqnarray*}
D_{H}^{2}(U,V) &=&\frac{\left\Vert \sqrt{\frac{dU}{d\lambda }}\right\Vert
^{2}+\left\Vert \sqrt{\frac{dV}{d\lambda }}\right\Vert ^{2}-2\left\langle
\sqrt{\frac{dU}{d\lambda }},\sqrt{\frac{dV}{d\lambda }}\right\rangle }{2} \\
&=&\frac{U(\mathcal{X})+V(\mathcal{X})}{2}-C_{B}(U,V).
\end{eqnarray*}
The intensity masses $U(\mathcal{X})$ and $V(\mathcal{X})$ are the expected
number of points of the respective Poisson point processes. Thus, Corollary~%
\ref{BDpro} provides another interesting interpretation: \emph{the
Bhattacharyya distance between two Poisson point processes is the difference
between the expected number of points per process and the Bhattacharyya
coefficient of their intensity measures}.

In general, the Hellinger distance cannot be numerically evaluated in closed
form. However, for Poisson point processes with Gaussian intensity function,
using the Bhattacharyya coefficient for Gaussians \cite{Kai67div}
\begin{multline*}
C_{B}\left( \mathcal{N}(\cdot ;\mu _{0},\Sigma _{0}),\mathcal{N}(\cdot ;\mu
_{1},\Sigma _{1})\right) =\\ \sqrt{(2\pi )^{d}\sqrt{\left\vert \Sigma _{0}\right\vert \left\vert \Sigma
_{1}\right\vert }}\mathcal{N}\left( \frac{\mu _{0}}{2};\frac{\mu _{1}}{2},%
\frac{\Sigma _{0}+\Sigma _{1}}{2}\right)
\end{multline*}%
yields an analytic expression for the Hellinger distance between the
Gaussian intensity functions, stated as follows.

\begin{corollary}
The Bhattacharyya distance between two Poisson point processes with Gaussian
intensities:
\begin{subequations}
\begin{align}
u(x)& =w_{u}\mathcal{N}(x;m_{u},P_{u}), \\
v(x)& =w_{v}\mathcal{N}(x;m_{v},P_{v})
\end{align}%
(measured in units of $K^{-1}$) is given by
\end{subequations}
\begin{multline}
D_{B}(F,G)=\frac{w_{u}+w_{v}}{2}- 
\sqrt{(2\pi)^dw_{u}w_{v}\sqrt{\left\vert P_{u}\right\vert \left\vert
P_{v}\right\vert }}\times\\ \mathcal{N}\left(\frac{m_{u}}{2};\frac{m_{v}}{2},\frac{%
P_{u}+P_{v}}{2}\right).
\end{multline}
\end{corollary}

\begin{remark}
For point processes, the Bhattacharyya distance can be defined by replacing
the standard (Lebesque) integral with the set integral. Again let $\phi $
and $\varphi $ denote the FISST densities of the respective point processes.
Then it follows from \cite{VSD05} that the corresponding probability
densities relative to $\mu $ are given by $f(X)=K^{|X|}\phi (X)$ and $%
g(X)=K^{|X|}\varphi (X)$. Hence,%
\begin{align*}
\left\langle \sqrt{f},\sqrt{g}\right\rangle _{\mu }& =\sum_{i=0}^{\infty }%
\frac{1}{i{!}K^{i}}\!\!\int_{\mathcal{X}^{i}}\!\sqrt{K^{i}\phi
(\{x_{1},...,x_{i}\}})\times\\
&\qquad\qquad\qquad\;\sqrt{K^{i}\varphi (\{x_{1},...,x_{i}\})}d(x_{1},...,x_{i}) \\
& =\int \sqrt{\phi (X)}\sqrt{\varphi (X)}\delta {X}
\end{align*}%
and the Bhattacharyya distance can be written in terms of FISST densities
and set integral as%
\begin{equation*}
D_{B}(\phi ,\varphi )=-\ln \int \sqrt{\phi (X)}\sqrt{\varphi (X)}\delta {X}.
\end{equation*}
\end{remark}

\section{Application to multi-target sensor management}

\label{sencon}

In this section, we present an application of our result to a sensor
management (a.k.a. sensor control) problem for multi-target systems, where system states are modeled as point processes or random finite sets (RFS) \cite{GMN97, MahlerPHD, VSD05, Mah07}. A multi-target system is fundamentally different from a
single-target system in that the number of states changes with time due to
births and deaths of targets.

For the purpose of illustrating the result in the previous section, we assume a linear Gaussian multi-target model \cite{VoMaGMPHD05}, where the hidden multi-target state at time $k$ is a finite set $X_{k}$, which is partially observed as another finite set $Z_{k}$. All aspects of the system dynamics as well as sensor detection and false alarms  are described in details in Appendix~\ref{App:model}. 

Multi-target sensor management is a stochastic control problem which involves the following steps
\begin{enumerate}
\item Propagating the multi-target posterior density, or alternatively a
tractable approximation, recursively in time;

\item At each time, determining the action of the sensor by optimizing an
objective function over a set of admissible actions.
\end{enumerate}

In step $1$, propagating the full posterior is generally intractable. However, for linear Gaussian multi-target systems, the first moment of the posterior (a.k.a. the intensity function) can be propagated efficiently via the Gaussian Mixture Probability Hypothesis Density (GM-PHD) filter \cite{VoMaGMPHD05} as documented in Appendix~\ref{App:propag}. The sensor action in step $2$ is executed by applying a control command/signal to the sensor, usually in order to either minimize a cost or maximize a reward. In the rest of this section, we demonstrate that the Cauchy-Schwarz divergence is a useful reward function for multi-target sensor management.

\subsection{Cauchy-Schwarz divergence based reward}
Denote by $\mathcal{R}(a_{k-1},Z_{k:k+p})$ the value of a reward function if the control command $a_{k-1}$ were applied to the sensor at time $k-1$ and subsequently the measurement sequence $Z_{k:k+p}=[Z_{k},Z_{k+1},...,Z_{k+p}]$ is observed for $p+1$ time steps in the future. For illustration purpose, we only focus on the single step look-ahead (i.e. $p=0$) policy. Naturally, given the reward function $\mathcal{R}(a_{k-1},Z_{k:k+p})$, the optimal control command $a_{k-1}^\ast$ is chosen to maximize the expected reward $\mathbb{E}\big[\mathcal{R}(a_{k-1},Z_{k})\big]$, where the expectation is taken over all possible values of the future measurement $Z_{k}$. A computationally cheaper approach is to maximize the ideal predicted reward $\mathcal{R}(a_{k-1},Z_{k}^{\ast })$ \cite{Mah04,RVC11,HV14sencon}, where $Z_{k}^{\ast }$ is the ideal predicted measurement from the predicted intensity $v_{k|k-1}$, that is, assuming no false alarms (zero clutter) and perfect target measurements (unity detection probability and negligible measurement noise). Other choices of objective functions are discussed in \cite{Mah04, RV10, RVC11,HV14sencon}.

A common class of reward functions for sensor control is that of information theoretic divergences between the predicted and posterior
probability densities. For example, in \cite{RV10, RVC11,
HV14sencon} the R\'{e}nyi divergence is employed to quantify the information gain from
the future measurements for a chosen control action. The main drawback of the R\'{e}nyi divergence
based approach is that it involves computation of integrals in infinite
dimensional spaces which is generally intractable.

As an alternative to the R\'{e}nyi divergence, we propose the use of the
Cauchy-Schwarz divergence for multi-target sensor control. According to
Proposition~\ref{CSpro}, computing the Cauchy-Schwarz reward function for Poisson multi-target densities
reduces to calculating the squared $L^{2}$-distance between the predicted
and posterior intensities:
\begin{equation}
\mathcal{R}(a_{k-1},Z_{k})=\frac{K}{2}\left\Vert v_{k|k-1}(\cdot
)-v_{k}(\cdot ;a_{k-1},Z_{k})\right\Vert ^{2}  \label{reward}
\end{equation}

This strategy effectively replaces the evaluation of the R\'{e}nyi
divergence, via integrals in the infinite dimensional space $\mathcal{F}(%
\mathcal{X}),$ with the Cauchy-Schwarz divergence, which can be computed via standard integrals on
the finite dimensional space $\mathcal{X}$. Moreover, when the GM-PHD filter
is used for the propagation of the Gaussian mixture posterior intensity, the
reward function $\mathcal{R}(a_{k-1},Z_{k})$ can be evaluated in closed
form using Corollary~\ref{CSdivGM}.

In this section, our control policy is to select the control command $a_{k-1}$ so as to maximize the ideal reward $\mathcal{R}(a_{k-1} ,Z^\ast_{k-1})$.
\subsection{Numerical example}

\label{sim}

\begin{figure}[tbh]
\centering
\includegraphics[scale=.55]{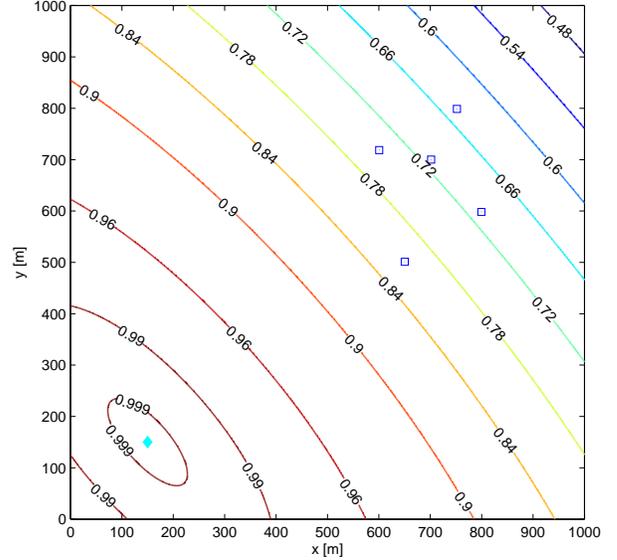}
\caption{Initial positions of the sensor ($\diamondsuit$) and targets ($\square$). The contours depict the sensor's
detection profile presented in \eqref{GMpd}, in which the detection probability decreases with distance from the sensor.}
\label{det_prof}
\end{figure}

This example is based on a scenario adapted from \cite{RVC11} in which a
mobile robot is tracking a varying number of moving targets. The
surveillance area is a square of dimensions $1000m\times 1000m$. Each target
at time $k-1$ is characterized by a single-target state of the form $x_{k-1}=[{p}_{k-1}^{T},{%
\dot{p}_{k-1}}^{T}]^{T}$ where ${p}_{k-1}$ is the 2D position vector and $\dot{p}%
_{k-1}$ is the 2D velocity vector. If the control command $a_{k-1}$ is applied at time $k-1$, the sensor will move from its current position $s_{k-1}$ to a new position ${s}_{k}(a_{k-1})$, where a target with state $x_{k}$ can be detected with probability
\begin{equation}\label{GMpd}
p_{D,k}({x}_{k};a_{k-1})=\frac{\mathcal{N}(s_{k}({a}_{k});Hx_{k},{S})}{%
\mathcal{N}(0;0,{S})}
\end{equation}%
where
\begin{equation*}
{H}=%
\begin{bmatrix}
1 & 0 & 0 & 0 \\
0 & 1 & 0 & 0%
\end{bmatrix}%
,\,{S}=10^{6}%
\begin{bmatrix}
3 & -2.4 \\
-2.4 & 3.6%
\end{bmatrix}%
.
\end{equation*}%
The detection profile is illustrated in Fig.~\ref{det_prof}.

The single-target transition density is $f({x}_{k}|{x}_{k-1})=\mathcal{N}({x}_{k};{F}{x}_{k-1},{Q})$, where
\begin{equation*}
{F}=%
\begin{bmatrix}
1 & 0 & T & 0 \\
0 & 1 & 0 & T \\
0 & 0 & 1 & 0 \\
0 & 0 & 0 & 1%
\end{bmatrix}%
,\;{Q}=27%
\begin{bmatrix}
T^{3} & 0 & \frac{T^{2}}{54} & 0 \\
0 & T^{3} & 0 & \frac{T^{2}}{54} \\
\frac{T^{2}}{54} & 0 & \frac{T}{81} & 0 \\
0 & \frac{T^{2}}{54} & 0 & \frac{T}{81}%
\end{bmatrix}%
\end{equation*}%
with $T=1s$.

Measurements are noisy position returns according to the single-target
likelihood \[g({z}_{k}|{x}_{k})=\mathcal{N}({z}_{k};{H}{x}_{k},{R}_{k}),\]
where
\begin{equation*}
{R}_{k}=\sigma _{\epsilon ,k}^{2}%
\begin{bmatrix}
1 & 0 \\
0 & 1%
\end{bmatrix}%
\end{equation*}
with $\sigma _{\epsilon ,k}=3m$.

Clutter is modeled by a Poisson RFS with intensity $\kappa ({z})=\lambda
c({z})$ where $\lambda =2\!\times\!10^{\!-\!5}m^{\!-\!2}$ and $c({z})=U([0,1000m]\times[0,1000m])$ is
the uniform density over the surveillance area.

At time $k-1$, the set $\mathbb{A}_{k-1}$ contains all admissible control
command that drive the sensor from the current position $s_{k-1}=\left[
s_{k-1}^{(x)},s_{k-1}^{(y)}\right] ^{T}$ to one of the following locations
\begin{equation*}
\mathbb{S}_{k}\!=\!\left\{\!\left[ s_{k\!-\!1}^{(x)}\!+\!j_{\!}\Delta _{R}\!\cos (_{\!}\ell \Delta
_{\theta\!}),s_{k\!-\!1}^{(y)}\!+\!j_{\!}\Delta _{R}\!\sin (_{\!}\ell \Delta _{\theta\!})_{\!}\right]
^{\!T\!}\right\}_{\!(j,\ell)=(0,0)}^{\!(N_{R},N_{\theta})},
\end{equation*}%
where $\Delta _{\theta }=\frac{2\pi}{N_{\theta}} {rad}$ and $\Delta _{R}=50 {m}$ are the
angular and radial step sizes respectively. The number of angular and radial
steps are $N_{R}=2$ and $N_{\theta }=8$. The set $\mathbb{S}_{k}$, thus, has $17$ options in total which
discretize the angular and radial region around the current sensor position. The sensor is always kept inside the surveillance area by setting the value of the objective function corresponding to positions outside the surveillance area to $-\infty $. 

\begin{figure}[h!]
\centering
\includegraphics[scale=0.53]{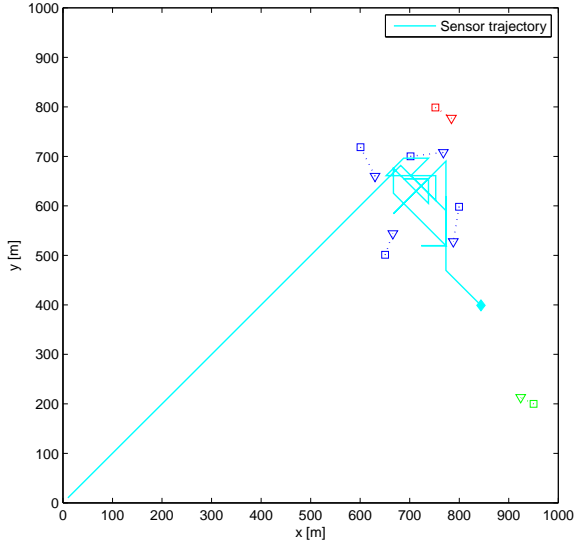}
\caption{A typical sensor trajectory. Target start and stop positions are marked by $\square$ and $%
\protect\nabla $, respectively. The red target died at $k=19$ whereas the
green target was born at $k=27$. The sensor initially moved towards the
targets and remained in their vicinity, then moved again to the middle of the existing targets and the new born target for optimal detection of all targets.}
\label{traj}
\end{figure}

With these settings, it is expected that our control policy should, intuitively speaking, move the
sensor towards the targets, and remain in their vicinity in order to obtain
a high detection probability. Fig.~\ref{traj}
depicts a typical sensor trajectory which appears to be consistent with
this intuitive expectation.

We proceed to illustrate the performance of the proposed strategy. First, we compare the performance of the Cauchy-Schwarz divergence based control strategy to that of an existing R\'{e}nyi divergence based control strategy proposed in \cite{RVC11}. Since the R\'{e}nyi divergence in general has no closed form solution and thus must be approximated by Sequential Monte Carlo (SMC), we also have to implement the Cauchy-Schwarz divergence using SMC approximation in order to enable a fair comparison. Second, the proposed GM implementation performance is then benchmarked against that of the SMC-based approach. When the objective function is approximated by SMC, the corresponding SMC-PHD filter \cite{VSD05} is used for recursive propagation of the posterior intensity function. All algorithms were implemented in MATLAB R2010b on a laptop with
an Intel Core i5-3360 CPU and 8GB of RAM. The average
run time for the R\'{e}nyi divergence based strategy is 10.62
seconds (SMC-PHD filter implementation) while those for the
Cauchy-Schwarz based strategies are 10.68 seconds (SMC-PHD filter implementation) and 3.21 seconds (GM-PHD filter
implementation). It is evident that the closed form Cauchy-Schwarz divergence based strategy is the fastest.

\begin{figure}[htb]
\centering
\includegraphics[scale=0.51]{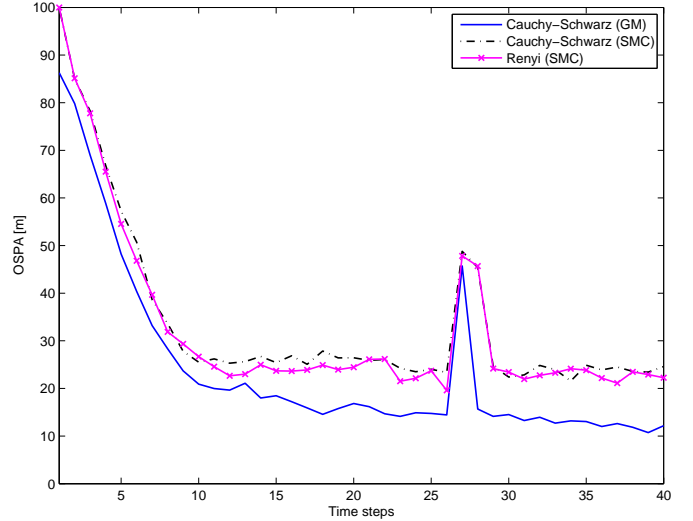}
\caption{Comparison of the averaged OSPA distance generated by different
control strategies. While SMC-PHD implementations for the R\'enyi divergence (dashed line) and the Cauchy-Schwarz divergence (starred line) yield similar results, they are outperformed by the GM-PHD implementation (solid line) due to closed form solution for the Cauchy-Schwarz divergence and better filtering performance.}
\label{OSPA}
\end{figure}

Fig.~\ref{OSPA} shows the Optimal SubPattern Assignment (OSPA) metric or
miss distance \cite{SVV08} (with parameters $p=2$, $c=100m$) averaged over
200 Monte Carlo runs for each of the considered control strategies. The OSPA
curves in Figure~\ref{OSPA} suggest that the closed form GM-PHD filter based
strategy outperforms its approximate SMC-PHD filter based counterparts,
while the performance of the two approximate SMC-PHD filter based strategies
are virtually identical.

These numerical results suggest that the Cauchy-Schwarz divergence can be at
least as effective as the R\'{e}nyi divergence when used as a reward
function for multi-target sensor control. The results further suggest that
the former has the distinct advantage of the GM implementation which leads to superior performance due to closed form solution and better filtering capability.

\section{Conclusions}

\label{sum}

In this paper, we have extended the notion of the Cauchy-Schwarz divergence
to point processes, and have shown that for an appropriate choice of
reference measure, the Cauchy-Schwarz divergence between the probability
densities of two Poisson point processes is half the squared distance
between their intensity functions. We have extended this result to mixtures
of Poisson point process and derived closed form expressions for the
Cauchy-Schwarz divergence when the intensity functions are Gaussian
mixtures. The Cauchy-Schwarz divergence for probability densities is not
necessarily invariant to the choice of reference measure. Nonetheless the
Cauchy-Schwarz divergence for the square roots of probability densities, or
equivalently, the Bhattacharyya distance for probability measures,
importantly is invariant to the choice of reference measure. For Poisson
point processes, our result implies that the Bhattacharyya distance between
the probability distributions is equal to the square of the Hellinger
distance between the intensity measures, which in turn is the difference
between the expected number of points per process and the Bhattacharyya
coefficient of their intensity measures. We have illustrated an application
of our result on a sensor control problem for multi-target tracking where
the system state is modeled as a point process. Our result is an addition to
the list of interesting properties of Poisson point processes and has
important implications in the approximation of point processes.

\appendices
\section{Linear Gaussian system model}
\label{App:model}

In a linear Gaussian multi-target model, each constituent element $x_{k-1}$ of the
multi-target state $X_{k-1}$ at time $k-1$ either continues to exist at
time $k$ with probability $p_{S,k}$ or dies with probability $1-p_{S,k}$, and
conditional on its existence at time $k$, transitions from $x_{k-1}$ to $%
x_{k}$ with probability density%
\begin{equation}
f(x_{k}|x_{k-1})=\mathcal{N}(x_{k};F_{k-1}x_{k-1},Q_{k-1}).
\end{equation}%
The surviving targets at time $k$ is thus a Multi-Bernoulli point process or
RFS \cite{Mah04, RV10, RVC11,HV14sencon}. New targets can arise at time $k$
either by spontaneous births, or by spawning from targets at time $k-1$. The
set of birth targets and spawned targets are modeled as Poisson point
processes with respective Gaussian mixture intensity functions
\begin{align*}
\gamma _{k}(x)& =\sum_{i=1}^{J_{\gamma ,k}}w_{\gamma ,k}^{(i)}\mathcal{N}%
\left( x;m_{\gamma ,k}^{(i)},P_{\gamma ,k}^{(i)}\right) , \\
\beta _{k|k-1}(x|\zeta )& =\sum_{i=1}^{J_{\beta ,k}}w_{\beta ,k}^{(i)}%
\mathcal{N}\left( x;F_{\beta ,k-1}^{(i)}\zeta +d_{\beta ,k-1}^{(i)},Q_{\beta
,k-1}^{(i)}\right) ,
\end{align*}%
The multi-target state is hidden and is partially observed by a sensor
driven by the control vector $a_{k-1}$ at time $k-1$. Each target evolves
and generates observations independently of one another. A target with state
$x_{k}$ is detected by the sensor with probability:
\begin{equation*}
p_{D,k}(x;a_{k-1})=\sum_{j=0}^{J_{D,k}}w_{D,k}^{(j)}\mathcal{N}\left(
x;m_{D,k}^{(j)}(a_{k-1}),P_{D,k}^{(j)}\right)
\end{equation*}
(or missed with probability $1-p_{D,k}(x_{k};a_{k-1})$) and conditional on
detection generates a measurement $z_{k}$ according to the probability
density%
\begin{equation}
g_{k}(z_{k}|x_{k})=\mathcal{N}(z_{k};H_{k}x_{k},R_{k}).
\end{equation}%
The detections corresponding to targets is thus a Multi-Bernoulli point
process \cite{Mah04, RV10, RVC11,HV14sencon}. The sensor also
registers a set of spurious measurements (clutter), independent of the
detections, modeled as a Poisson point process with intensity $\kappa_{k}$. Thus, at each time step the measurement is a collection of detections $Z_{k}$, only some of which are generated by targets.

\section{Posterior intensity propagation}\label{App:propag}
In general, posterior intensity function is propagated recursively in time via the Probability Hypothesis Density (PHD) filter \cite{MahlerPHD}. For the linear Gaussian multi-target system described in Appendix~\ref{App:model}, the posterior intensity function is propagated via the Gaussian Mixture PHD (GM-PHD) filter \cite{VoMaGMPHD05} as follows\footnote{Here, we use a slightly different technique from that in \cite{UlmkeCPHD10}, which proposes an approximate propagation for the original GM-PHD filter in order to mitigate computational issues involving negative Gaussian mixture weights which arise due to a state dependent detection probability. For notational compactness we omit the time
index on the state variable and the conditioning on the measurement history
in expressions involving the posterior intensity function.}.

\emph{Prediction:} If the posterior intensity at time $k-1$ is a Gaussian
mixture of the form
\begin{equation*}
v_{k-1}(x)=\sum_{i=1}^{J_{k-1}}w_{k-1}^{(i)}\mathcal{N}%
(x;m_{k-1}^{(i)},P_{k-1}^{(i)})
\end{equation*}%
then the predicted intensity at time $k$ is also a Gaussian mixture and is
given by
\begin{equation*}
v_{k|k-1}(x)=v_{S,k|k-1}(x)+v_{\beta ,k|k-1}(x)+\gamma _{k}(x)
\end{equation*}%
where
\begin{align*}
v_{S,k|k-1}(x)& =p_{S,k}\sum_{i=1}^{J_{k-1}}w_{k-1}^{(i)}\mathcal{N}\left(
x;m_{S,k|k-1}^{(i)},P_{S,k|k-1}^{(i)}\right)  \\
v_{\beta ,k|k-1}(x)& =\sum_{i=1}^{J_{k-1}}\sum_{j=1}^{J_{\beta
,k}}w_{k-1}^{(i)}w_{\beta ,k}^{(j)}\mathcal{N}\left( x;m_{\beta
,k|k-1}^{(i,j)},P_{\beta ,k|k-1}^{(i,j)}\right)  \\
m_{S,k|k-1}^{(i)}& =F_{k-1}m_{k-1}^{(i)} \\
P_{S,k|k-1}^{(i)}& =Q_{k-1}+F_{k-1}P_{k-1}^{(i)}\left[ F_{k-1}\right] ^{T} \\
m_{\beta ,k|k-1}^{(i,j)}& =F_{\beta ,k-1}^{(j)}m_{k-1}^{(i)}+d_{\beta
,k-1}^{(j)} \\
P_{\beta ,k|-1}^{(i,j)}& =Q_{\beta ,k-1}^{(j)}+F_{\beta ,k-1}^{(j)}P_{\beta
,k-1}^{(i)}\left[ F_{\beta ,k-1}^{(j)}\right] ^{T}.
\end{align*}

\emph{Update:} If predicted intensity and detection probability are Gaussian
mixtures of the form
\begin{equation*}
v_{k|k-1}(x)=\sum_{i=1}^{J_{k|k-1}}w_{k|k-1}^{(i)}\mathcal{N}\left(
x;m_{k|k-1}^{(i)},P_{k|k-1}^{(i)}\right), 
\end{equation*}%
then, the posterior intensity at time $k$ is given by
\begin{equation*}
v_{k}(x;Z_{k}(a_{k-1}))=v_{M,k}(x;a_{k-1})+\sum_{z\in
Z_{k}(a_{k-1})}v_{D,k}(x;z)
\end{equation*}%
where
\begin{align*}
v_{M,k}(x;a_{k-1})& =\sum_{i=1}^{J_{k|k-1}}w_{M,k}^{(i)}(a_{k-1})\mathcal{N}%
\left( x;m_{k|k-1}^{(i)},P_{k|k-1}^{(i)}\right)  \\
w_{M,k}^{(i)}(a_{k-1})& =\frac{w_{\mu ,k}^{(i)}(a_{k-1})T_{k}(a_{k-1})}{%
\displaystyle\sum_{i=1}^{J_{k|k-1}}w_{\mu ,k}^{(i)}(a_{k-1})} \\
w_{\mu ,k}^{(i)}(a_{k-1})& =\left[ 1-p_{D,k}\left(
m_{k|k-1}^{(i)};a_{k-1}\right) \right] w_{k|k-1}^{(i)} \\
T_{k}(a_{k-1})&
=\sum_{i=1}^{J_{k|k-1}}w_{k|k-1}^{(i)}-\sum_{i=1}^{J_{k|k-1}}%
\sum_{j=0}^{J_{D,k}}w_{k|k-1}^{(i,j)}(a_{k-1}) \\
w_{k|k-1}^{(i,j)}(a_{k-1})&
=w_{k|k-1}^{(i)}w_{D,k}^{(j)}q_{k|k-1}^{(i,j)}(a_{k-1}), \\
q_{k|k-1}^{(i,j)}(a_{k-1})& =\mathcal{N}\left(
m_{D,k}^{(j)}(a_{k-1});m_{k|k-1}^{(i)},P_{k|k-1}^{(i)}+P_{D,k}^{(j)}\right) ,
\end{align*}%
and
\begin{align*}
v_{D,k}(x;z)& =\sum_{i=1}^{J_{k|k-1}}\sum_{j=0}^{J_{D,k}}w_{k}^{(i,j)}(z)%
\mathcal{N}\left( x;m_{k|k}^{(i,j)}(z),P_{k|k}^{(i,j)}\right) , \\
w_{k}^{(i,j)}(z)& =\frac{w_{k|k-1}^{(i,j)}(a_{k-1})q_{k}^{(i,j)}(z)}{\kappa
_{k}(z)+\displaystyle\sum_{i=1}^{J_{k|k-1}}%
\sum_{j=0}^{J_{D,k}}w_{k|k-1}^{(i,j)}(a_{k-1})q_{k}^{(i,j)}(z)}, \\
q_{k}^{(i,j)}(z)& =\mathcal{N}\left(
z;H_{k}m_{k|k-1}^{(i,j)},R_{k}+H_{k}P_{k|k-1}^{(i,j)}H_{k}^{T}\right) , \\
m_{k|k-1}^{(i,j)}& =m_{k|k-1}^{(i)}+K_{k|k-1}^{(i,j)}\left[
m_{D,k}^{(j)}(a_{k-1})-m_{k|k-1}^{(i)}\right] , \\
P_{k|k-1}^{(i,j)}& =\left[ I-K_{k|k-1}^{(i,j)}\right] P_{k|k-1}^{(i)}, \\
K_{k|k-1}^{(i,j)}& =P_{k|k-1}^{(i)}\left[ P_{k|k-1}^{(i)}+P_{D,k}^{(j)}%
\right] ^{-1}, \\
m_{k|k}^{(i,j)}(z)& =m_{k|k-1}^{(i,j)}+K_{k}^{(i,j)}\left[
z-H_{k}m_{k|k-1}^{(i,j)}\right] , \\
P_{k|k}^{(i,j)}& =\left[ I-K_{k}^{(i,j)}H_{k}\right] P_{k|k-1}^{(i,j)}, \\
K_{k}^{(i,j)}& =P_{k|k-1}^{(i,j)}H_{k}^{T}\left(
H_{k}P_{k|k-1}^{(i,j)}H_{k}^{T}+R_{k}\right) ^{-1},
\end{align*}%
and by convention $q_{k|k-1}^{(i,0)}=1$, $m_{k|k-1}^{(i,0)}=m_{k|k-1}^{(i)}
$, and $P_{k|k-1}^{(i,0)}=P_{k|k-1}^{(i)}$.

%



%
\begin{IEEEbiographynophoto}{}
\end{IEEEbiographynophoto}
\begin{IEEEbiographynophoto}{}
\end{IEEEbiographynophoto}
\begin{IEEEbiographynophoto}
{Hung Gia Hoang} was born in Ha Noi, Viet Nam, in 1978. He received the B.E. degree in Electronic Engineering and Telecommunications in 2000 from Vietnam National University (VNU Hanoi) and the Ph.D. degree in Electrical Engineering in 2009 from the University of New South Wales, Australia. He is currently a Research Fellow in the Department of Electrical and Computer Engineering at Curtin University. His research interests are probabilistic methods in signal processing, control, and information theory.
\end{IEEEbiographynophoto}
\enlargethispage{-3.84in}
\begin{IEEEbiographynophoto}
{Ba-Ngu Vo} received his Bachelor degrees jointly in Science and Electrical Engineering with first class honors in 1994, and PhD in 1997. He had held various research positions before joining the department of Electrical and Electronic Engineering at the University of Melbourne in 2000. In 2010, he joined the School of Electrical Electronic and Computer Engineering at the University of Western Australia as Winthrop Professor and Chair of Signal Processing. Since 2012 he is Professor and Chair of Signals and Systems in  the Department of Electrical and Computer Engineering at Curtin University.  Prof. Vo is a recipient of the Australian Research Council's inaugural Future Fellowship and the 2010 Australian Museum Eureka Prize for Outstanding Science in support of Defence or National Security. His research interests are Signal Processing, Systems Theory and Stochastic Geometry with emphasis on target tracking, robotics, computer vision and space situational awareness.
\end{IEEEbiographynophoto}

\begin{IEEEbiographynophoto}
{Ba-Tuong Vo} was born in Perth, Australia, in 1982. He received the B.Sc. degree in applied mathematics and B.E. degree in electrical and electronic engineering (with first-class honors) in 2004 and the Ph.D. degree in engineering (with Distinction) in 2008, all from the University of Western Australia. He is currently an Associate Professor in the Department of Electrical and Computer Engineering at Curtin University and a recipient of an Australian Research Council Fellowship. His primary research interests are in point process theory, filtering and estimation, and multiple object filtering. Dr. Vo is a recipient of the 2010 Australian Museum DSTO Eureka Prize for "Outstanding Science in Support of Defence or National Security".
\end{IEEEbiographynophoto}

\begin{IEEEbiographynophoto}
{Ronald Mahler} Ronald Mahler was born in Great Falls, MT, in 1948.  He received the B.A. degree in mathematics from the University of Chicago, Chicago, IL, in 1970; the Ph.D. in mathematics from Brandeis University, Waltham, MA, in 1974; and the B.E.E. in electrical engineering from the University of Minnesota, Minneapolis, in 1980.  He was an Assistant Professor of Mathematics at the University of Minnesota from 1974 to 1979.  He was employed at Lockheed Martin Corp. from 1980 through 2014, where he received the 2004 and 2008 MS2 Author of the Year awards.  Currently he is an independent consultant (Random Sets LLC).  His research interests include information fusion, expert systems theory, multitarget tracking, and sensor management.  He has authored two books and coauthored a third.  Two of his papers are the first and fourth most-cited papers published in \textit{IEEE Trans. Aerospace \& Electronic Systems} over the last decade.  He has received the 2007 Mignogna Data Fusion Award, the 2005 IEEE AESS Harry Rowe Mimno Award, and the 2007 IEEE AESS M. Barry Carlton Award.
\end{IEEEbiographynophoto}
\enlargethispage{-3.81in}







\end{document}